\documentclass[prx,aps,reprint,superscriptaddress,preprintnumbers,
longbibliography
]{revtex4-1}
\usepackage{graphicx}
\usepackage{dcolumn}
\usepackage{bm}
\usepackage{color}
\usepackage{sidecap}
\usepackage{amssymb}

\begin{document}

\title {Coexistence of high-bit-rate quantum key distribution and data on optical fiber}
\author{K. A. Patel}
\affiliation{Toshiba Research Europe Limited, Cambridge Research Laboratory, 208 Cambridge Science Park, Milton Road, Cambridge, CB4~0GZ, United Kingdom}
\affiliation{Cambridge University Engineering Department, 9 J J Thomson Avenue, Cambridge, CB3 0FA, United Kingdom}
\author {J. F. Dynes}
\affiliation{Toshiba Research Europe Limited, Cambridge Research Laboratory, 208 Cambridge Science Park, Milton Road, Cambridge, CB4~0GZ, United Kingdom}
\author {I. Choi}
\author{A. W. Sharpe}
\affiliation{Toshiba Research Europe Limited, Cambridge Research Laboratory, 208 Cambridge Science Park, Milton Road, Cambridge, CB4~0GZ, United Kingdom}
\author{A.~R.~Dixon}
\author {Z.~L.~Yuan}
\email{zhiliang.yuan@crl.toshiba.co.uk}
\affiliation{Toshiba Research Europe Limited, Cambridge Research Laboratory, 208 Cambridge Science Park, Milton Road, Cambridge, CB4~0GZ, United Kingdom}
\author{R.~V.~Penty}
\affiliation{Cambridge University Engineering Department, 9 J J Thomson Avenue, Cambridge, CB3 0FA, United Kingdom}
\author {A.~J.~Shields}
\email{andrew.shields@crl.toshiba.co.uk}
\affiliation{Toshiba Research Europe Limited, Cambridge Research Laboratory, 208 Cambridge Science Park, Milton Road, Cambridge, CB4~0GZ, United Kingdom}
\date{\today}

\begin{abstract}
Quantum key distribution (QKD) uniquely allows distribution of cryptographic keys with security verified by quantum mechanical limits.
Both protocol execution and subsequent applications require \textcolor{black}{the} assistance of classical data communication channels. While using separate fibers is one option, it is economically more viable if data and quantum signals are simultaneously transmitted through a single fiber. However, noise-photon contamination arising from the intense data signal has severely restricted both the QKD distances and secure key rates.
Here, we exploit a novel temporal-filtering effect for noise-photon rejection.
This allows high-bit-rate QKD over fibers up to 90~km in length and populated with error-free bidirectional Gb/s data communications.  With high-bit rate and range sufficient for important information infrastructures, such as smart cities and 10~Gbit Ethernet,  QKD is a significant step closer towards wide-scale deployment in fiber networks.
\end{abstract}

\maketitle

\noindent 
Quantum key distribution (QKD)\cite{bb84,ekert91} has been established as a viable technology over dedicated fibers \cite{elliot05,peev09,sasaki11}.
In the absence of data signals on the same fiber,
secure key rates exceeding 1~Mb/s \cite{dixon08,zhang09,dixon10} and \textcolor{black}{a transmission distance of over 250~km} \cite{stucki09b,wang12} have been achieved. To date, most experiments and field trials have been performed on dark fibers. As dark fiber is a scarce and expensive resource, there is a pressing need \textcolor{black}{to enable} QKD's coexistence with data signals on the same fiber \cite{townsend97,chapuran09,choi11,eraerds10,lancho10,qi10}. However, all work so far (see Table~\ref{tab:comparison}) has been limited to very low bit rates, short fiber spans, and/or unidirectional data communications. Using a novel temporal-filtering effect, we demonstrate QKD in the presence of error-free bidirectional Gb/s data transfer with a secure bit rate which is over three orders of magnitude higher than previously reported.

The main challenge for the coexistence of quantum and data signals on the same fiber arises from the extreme contrast in their intensities.
Each quantum signal typically contains approximately 0.5 photons per pulse when implementing decoy protocols with weak laser pulses\cite{lo05,wang05}, while a data-laser pulse may contain 10$^{6}$ photons or more for a Gb/s link.
Although the data-laser signal can be readily filtered using wavelength multiplexing, secondary photons, resulting from its Raman and nonlinear interaction with the fiber, are impossible to reject completely because of their spectral overlap with the quantum signal.
Placing the quantum channel spectrally far away from the data channels can reduce the spectral overlap. However,in such systems the quantum channel is often in the 1310~nm band \cite{chapuran09,townsend97,choi11} or shorter ($<1~\mu$m) \cite{fernandez07,holloway11}. The fiber transmission loss is much higher at these wavelengths, which further restricts the QKD distance and secure key rate.  Other common techniques for noise-photon rejection include reducing the data-laser intensities \cite{townsend97,eraerds10} and spectral filtering \cite{eraerds10}. Exploiting data pulse gaps has also been demonstrated to suppress Raman photons scattered from copropagating data pulses \cite{choi11}.

\begin{table*}
\caption{Summary of existing quantum/data multiplexing demonstrations.}
\begin{tabular}{l|l|l|l|l}
  \hline \hline
  &QKD wavelength (nm) & Data wavelength (nm) & Distance (km) & Bit rate (kbit/s)\\ 
  \hline
  \textbf{BT}\cite{townsend97} & 1310 & 1550 & 28 & - \\
  \hline
  \textbf{Telcordia}\cite{chapuran09}& 1310 & 1550 & 25 & 0.006 \\
  \hline
  \textbf{Cork}\cite{choi11} & 1310 & 1290, 1550 & 10 & 1.3 \\
  \hline
  \textbf{Geneva}\cite{eraerds10} & 1551.72 & 1555.33--1555.75& 50 & 0.011 \\
  \hline
  \textbf{Madrid}\cite{lancho10} & 1550 & 1310, 1490 & 10 & 0.1 \\
  \hline\hline
  \textbf{This work} & 1550 & 1591--1611 & 50      &  507 \\
                     &      &            & 90      & 7.6     \\
  \hline \hline
  \end{tabular}
  \label{tab:comparison}
\end{table*}

Raman photons reach the detector at random times with respect to the regularly pulsed quantum signals. We show that this randomness can be exploited for enhancing the quantum signal to Raman noise ratio (SNR). Using subnanosecond gated InGaAs avalanche photodiodes (APDs) \cite{yuan07}, we have achieved a ten-fold enhancement in the SNR through temporal-filtering, thereby demonstrating high-bit-rate QKD over record distances of a single fiber multiplexed with 1~Gb/s error-free bidirectional data signals.

\begin{figure}[b]
\centering
\includegraphics[width=.96\columnwidth]{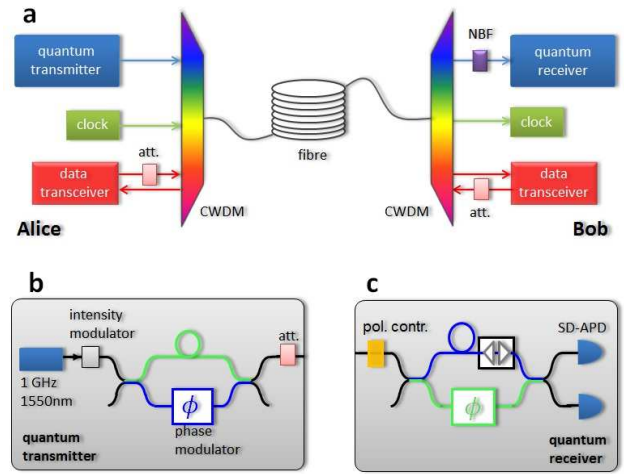}
\caption{Experimental setup. (a) Schematics for multiplexing of quantum, clock and data channels; (b) Quantum transmitter; (c) Quantum receiver. SD-APD: self-differencing avalanche photodiodes; att.: optical attenuator; CWDM: coarse wavelength division multiplexer; NBF: narrow bandpass filter (0.56~nm).
}
\label{fig:setup}
\newpage
\end{figure}

Figure~\ref{fig:setup}(a) shows the experimental setup. Two communicating parties, referred to as Alice and Bob, are linked by a single fiber. At each party, there are three subsystems for quantum, clock and data communications.  The quantum subsystem is described in Appendix A.
Both quantum and clock channels are unidirectional from Alice to Bob, while the data channel is formed from a symmetric bidirectional Gb/s link running \textcolor{black}{at the standard data clock rate of 1.25~Gb/s}. These channels are multiplexed using coarse wavelength division multiplexers (CWDM) for transmission through the single fiber.
The CWDMs feature an insertion loss of 0.5---1~dB at passbands, centered at 1551, 1571, 1591 and 1611~nm.  The fiber link is made of dispersion-shifted fiber featuring low chromatic dispersion of 4~ps/nm$\cdot$km and a measured loss of 0.2~dB/km at 1550~nm. Standard single-mode fiber can also be used by precompensating fiber dispersion at longer distances (greater than 50~km) \cite{yuan08}. To minimise loss of quantum transmission, the 1551~nm CWDM band is assigned to the quantum subsystem.

\begin{figure}[b]
\newpage
\centering
\includegraphics[width=1\columnwidth]{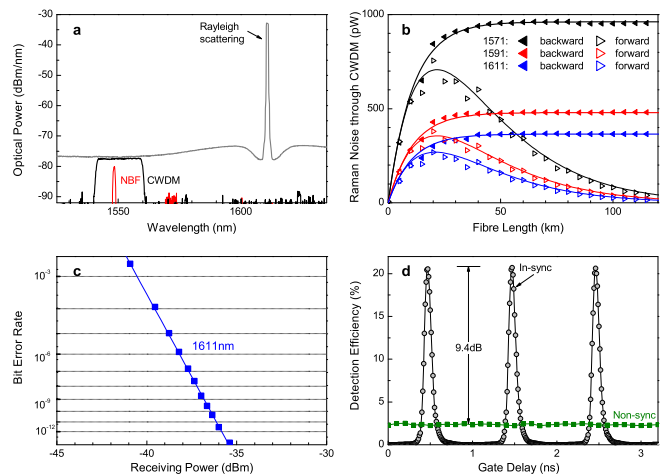}
\caption{Raman noise and its rejection. (a) Spectrum of back-scattered Raman noise measured over 80~km with a 0~dBm launch laser at a wavelength of 1611~nm; also shown are spectra after a CWDM filter only, and a combination of a CWDM coupler and a narrow band pass (NBF) filter. (b) Measured (symbols) and calculated (solid lines) Raman noise power into the quantum receiver through Bob's CWDM coupler; (c) Bit error rate measured over a fiber link of 80~km for the 1611~nm receiver as a function of receiving power; (d) Single-photon detection efficiencies  as a function of gate delay of the gated detector under synchronized (circles) and non-synchronized (squares) illuminations. The illumination source is a pulsed laser clocked at 1~GHz with an average flux of 0.02~photons per pulse. }
\label{fig:raman}
\end{figure}

Figure~\ref{fig:raman}(a) shows a spectrum of back-scattered secondary photons generated by a 1611~nm \textcolor{black}{continuous-wave} laser \textcolor{black}{(linewidth: $<$0.1~nm)} launched into an 80~km fiber at 0~dBm power (1~mW). Rayleigh scattered photons are approximately 4 orders of magnitude more intense than the Raman-scattered photons. As the Rayleigh photons have the same wavelength as the 1611~nm laser, they can be readily rejected from the 1551~nm CWDM passband used by the quantum subsystem, as shown in Fig.~\ref{fig:raman}(a). However, Raman photons spectrally extend over the 1551~nm passband. Consequently, a considerable fraction of Raman photons enter the quantum receiver through the CWDM coupler.

We have systematically studied how much light is Raman-scattered into the 1551~nm passband by other CWDM channels in order to assign the wavelengths for classical communication. A 1~mW  \textcolor{black}{continuous-wave} laser signal is launched into the fiber link through one of the remaining CWDM channels at either Alice's or Bob's side, and we measure the scattered light power in the 1551~nm output of Bob's CWDM module. The measured power quantifies the amount of Raman scatter entering the quantum receiver through the 1551~nm channel. Figure~\ref{fig:raman}(b) shows the Raman scatter power (symbols) as a function of fiber length between Alice and Bob in 5~km intervals for three different CWDM channels. For each channel the backward scatter (light launched on Bob's side) and forward scatter (light launched on Alice's side) are shown, together with the result of a theoretical calculation (solid lines). This calculation is outlined in Appendix C.

Forward and backward scatter display a distinctively different behavior with increasing fiber length. Whereas forward scatter reaches a maximum value at a distance of about 20~km before it starts to decline, backward scatter saturates and does not decrease with distance. In the case of forward scatter the accumulation of Raman power along the fiber is eventually outstripped by the increasing fiber attenuation, leading to a reduction of Raman noise. In contrast, backward scatter travels back to the quantum receiver and is not subjected to higher loss with increasing distance. Hence, backward scatter never decreases but reaches saturation asymptotically.

At each fixed wavelength, the backward Raman scatter is always stronger than the forward scatter, and becomes dominant for long fibers. Additionally, for all wavelengths studied, the further the laser is spectrally away from the 1511~nm passband, the weaker the Raman scatter. Assigning Bob's data-laser to the 1611nm channel therefore minimizes the Raman scatter into the quantum receiver, as this configuration minimizes the amount of back scatter.  The two remaining wavelengths of 1571 and 1591~nm are assigned to Alice's lasers. As the clock laser needs comparably lower launch power (see Appendix B), it is preferable to assign the shorter wavelength of 1571~nm to the clock subsystem.

\begin{figure}[t]
\newpage
\centering
\includegraphics[width=.96\columnwidth]{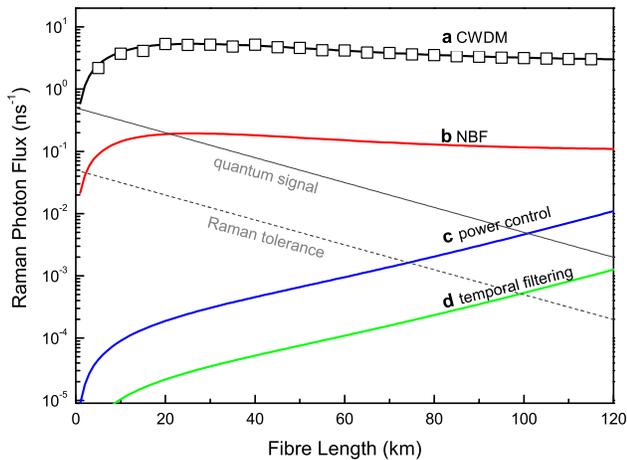}
\caption{Raman noise into the quantum receiver. (a) Measured (symbol) and calculated (solid line) Raman noise after Bob's CWDM; (b) Raman noise after the narrow band pass filter (NBF);  In both (a) and (b), two 0~dBm data-lasers of 1591 and 1611~nm are launched simultaneously at Alice and Bob, respectively.
(c) Reduced Raman noise after lowering the laser launch powers; (d) Effective Raman noise received within the active time of the gated detectors. Solid lines (b--c) are calculated results.}
\label{fig:noise_power}
\end{figure}

Figure~\ref{fig:noise_power} compares the combined Raman noise caused by both Alice's (1591~nm) and Bob's (1611~nm) data-lasers with the strength of the quantum signal at a flux of 0.5 photons per pulse at 1~GHz (thin solid line).
We omit the contribution of the clock laser because of its low launch power (see Appendix B).  The Raman noise is stronger than the quantum signal for every fiber length, especially for long fibers. At 90~km, the Raman noise is approximately 27~dB stronger than the quantum signal. Such a noise level would result in a quantum bit error rate (QBER) close to 50\%, preventing formation of a secure key.
To obtain a secure key, the QBER must be below 10\%, a typical threshold value for the decoy-state BB84 protocol \cite{lo05,wang05}.
Considering that other noise sources, such as encoding apparatus imperfections, detector dark counts and afterpulsing, may contribute around 5\% to the QBER, the Raman noise needs to be 10~dB weaker than the quantum signal as a practical guideline.  We refer to this level as the Raman tolerance, as plotted in Fig.~\ref{fig:noise_power}.

In addition to temporal-filtering,  we employ conventional techniques for the suppression of Raman noise.
As the first step, we place a narrow bandpass filter (NBF) in front of the quantum receiver, as shown in Fig.~\ref{fig:setup}(a). The filter has a passband of 0.56~nm, see Fig.~\ref{fig:raman}(a). Including its intrinsic loss of 0.6~dB, the filter reduces the Raman noise by 15~dB. The overall improvement in the SNR is 14.4~dB. Despite the improvement,  the Raman noise remains considerably stronger than the tolerance for most fiber lengths, as shown in Fig.~\ref{fig:noise_power}(b). QKD is possible only over very short lengths (approximately 3~km).

The next step is to lower the launch power of the data-lasers using optical attenuators (Fig.~\ref{fig:setup}(a)) to match the sensitivity of the data photo-receivers. As an example, Fig.~\ref{fig:raman}(c) shows the bit error ratio as a function of receiving power for the 1611~nm data channel. Its sensitivity, defined as the minimum receiving power required to achieve a bit error ratio no higher than $10^{-9}$, is measured to be -36.8~dBm at a data modulation rate of 1.25~Gb/s over a fiber link of 80~km.
Taking the fiber loss (0.2~dB/km) into account, a launch power much lower than 0~dBm can be used to achieve error-free data communications. For example, a launch power of -18.5~dBm is more than sufficient for 80~km data transmission.
With lower launch powers, the Raman noise (Fig.~\ref{fig:noise_power}(c)) is reduced considerably.

Applying temporal filtering, a new and crucial technique, to the conventional toolbox discussed above for noise
reduction, we can now reduce the Raman noise further to below the Raman-tolerance threshold for distances up to
100~km. Here, we operate InGaAs APDs using an alternating bias with a repetition frequency of 1~GHz.
\textcolor{black}{With a passive circuit for detection of extremely weak avalanches, the detector has been demonstrated to have an ultrashort dead time of less than 2~ns and support high count rates \cite{yuan07,dixon09,patel12}. The photon detection efficiency is independent of the incident photon flux, which is an underlying assumption required in the decoy-state QKD protocol.}
Figure~\ref{fig:raman}(d) shows the detection efficiency as a function of the detector gate delay under pulsed laser excitation. When the detector and laser are synchronized, the detector exhibits a peak detection efficiency of 20\%.
In contrast, after delaying the detection gate relative to the laser by 100~ps, the detection efficiency drops sharply to virtually zero.
The full width at the half maximum for each efficiency peak is measured as 100~ps, which is much shorter than the nominal detection window of 500~ps. This is due to the low-noise evolution of avalanches \cite{yuan10c}: Only avalanches triggered at the front edge of each gate can grow sufficiently strong to be detected.

The short active time of 100~ps reduces the impact of the Raman noise on QKD remarkably. The detector is effectively a temporal filter, rejecting those photons arriving outside of the active times.
The random arrival time of Raman photons is simulated by breaking the synchronization between the pulsed laser and detector. As shown in Fig.~\ref{fig:raman}(d), the detection efficiency for these randomly arriving photons is now reduced to approximately 2\%, which is almost 10 times lower than the peak efficiency for synchronized photons. The efficiency contrast results in a temporal rejection of 9.4~dB for the Raman photons. Now, the calculation shows the Raman noise is tolerable for fiber distances up to 100~km [Fig.~\ref{fig:noise_power}(d)].

\begin{figure}[t]
\newpage
\centering
\includegraphics[width=.88\columnwidth]{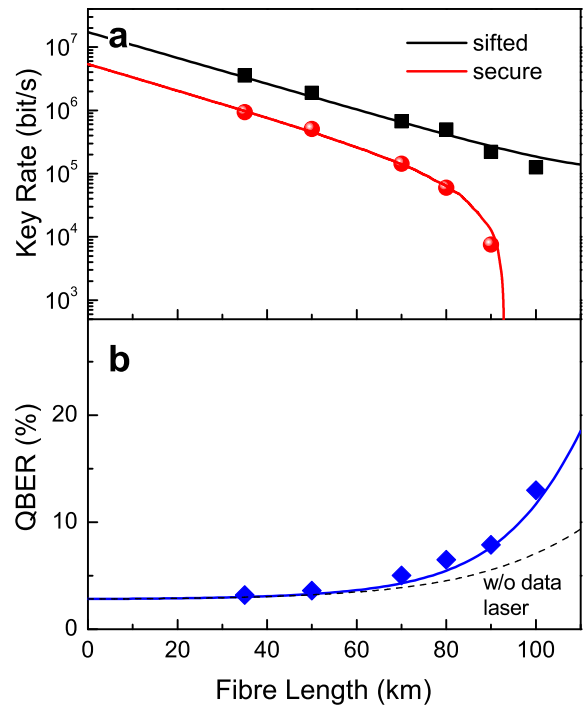}
\caption{QKD performance with error-free bidirectional Gb/s data channels. (a) Calculation (line) and measurement (symbols) of sifted and secure key rates as a function of fiber length; (b) Calculation (line) and measurement (symbols) of QBER. Also shown is the calculation of the QBER without contribution from data-lasers (dashed line).}
\label{fig:performance}
\newpage
\end{figure}

We performed QKD experiments using a single fiber shared simultaneously with optical clock synchronization (see Appendix B) and bidirectional error-free 1.25~Gb/s data communication. In the continuously operating quantum subsystem, the decoy-state BB84 protocol is implemented with three different pulse intensities.
We determine the secure key rate from Koashi's security proof \cite{koashi06}, following the approach of Rice and Harrington \cite{rice09} to estimate single-photon parameters from decoy states.
Figure~\ref{fig:performance}(a) plots the sifted and secure bit rates as a function of fiber length.   The sifted key rate falls off exponentially with fiber length at a rate of approximately 0.20~dB/km, which is the characteristic loss of the fiber. The secure key rate decreases at the same rate for short fiber distances (less than 50~km). We determine the secure key rate as 935 and 507 kbit/s over 35~km and 50~km fibers, respectively. Increasing the fiber length further, the secure bit rate decreases at a rate noticeably faster than the fiber loss, due to the increased cost of privacy amplification for higher \textcolor{black}{QBERs}. At 80 and 90~km, the secure rates are determined to be 72 and 7.6 kbit/s, respectively.

Figure~\ref{fig:performance}(b) shows the measured QBER (symbol) as a function of fiber length.  Detector afterpulsing and apparatus imperfection make up a floor of 3\% for short fiber lengths (less than 50~km). At these distances, both detector dark counts and Raman noise are negligible.
For fiber lengths greater than 50~km, the QBER increases gradually, because the dark counts and Raman contribution are no longer negligible as compared with the signal counts. To illustrate the contribution from the Raman noise, we plot the simulation of the QBER without data-lasers (dashed line) in Fig.~\ref{fig:performance}(b). At 90~km, the dark counts contribute 2.5\% and the Raman noise contributes 2.4\% towards the total QBER of 7.9\%.  At 100~km, the measured QBER exceeds 10\%, and hence no secure keys can be formed.

Using experimentally measured parameters only, we simulate the secure key rates and QBER, as shown by the solid lines in Figs.~\ref{fig:performance}(a) and (b). The simulation process is described in detail in Appendix C. We integrate both the forward and backward Raman scatters, and apply 9.4~dB temporal-filtering into the simulation. We also take into account extra loss due to fiber connectors and fiber dispersion. At 90~km, the connectors make up an additional 0.6~dB loss while the fiber dispersion adds 1~dB penalty to the data channels. The simulation is in excellent agreement with the experimental results.

In comparison with previous demonstrations (Table~\ref{tab:comparison}), the present work has achieved not only a much longer fiber span but also orders of magnitude higher secure key rates.
We believe this advance will have significant impact for future deployment of QKD technology and networks.
Firstly, the demonstrated distance of 90~km exceeds the optimal span for a topologically-optimized quantum network \cite{alleaume09}, and is longer than all the links demonstrated in quantum networks to date \cite{elliot05,peev09,sasaki11}.
Secondly, the reach distance is sufficiently to serve most links \textcolor{black}{in metropolitan networks \cite{ituG984}}. In  particular, it is sufficient to support smart cities, where a typical link spans from 30 -- 80~km \cite{smartcity}. Thirdly, the QKD system is capable of supporting 10~Gb Ethernet traffic, which is important for low cost implementation, reliability and straightforward installation and maintenance.
With 10Gb/s data channels, the reach distance will be reduced to 65~km, due to the lower receiver sensitivity at this data rate \cite{TenG}.  Nevertheless, this reach distance exceeds 40~km, the maximum fiber length defined in one 10~Gb Ethernet standard \cite{10GbEthernet}. With ability to support 10Gb/s Ethernet, QKD will be able to seamlessly integrate into important information infrastructures, such as business continuance and disaster recovery, distributed storage networks, and remote back-up, to offer the strongest cryptographic protection.

To conclude, we have shown the coexistence of QKD and Gb/s data communications over a single fiber up to 90~km. In achieving this, the Raman noise has been strongly suppressed by wavelength and temporal-filtering. Following this breakthrough on communication range and bit rate, we expect QKD will be an attractive resource for securing data communication networks.

\section*{Appendix}

\subsection{Quantum subsystem.}
Figures~\ref{fig:setup}(b) and (c) show the optical layout of the quantum transmitter and receiver. The transmitter consists of a 1550~nm pulsed laser, an intensity modulator, an asymmetric Mach-Zehnder interferometer and an optical attenuator. The receiver consists of a \textcolor{black}{polarization} controller, an asymmetric Mach-Zehnder interferometer that matches the one in the transmitter, and two self-differencing (SD) single-photon detectors.

The quantum system implements the standard BB84 protocol with decoy states \cite{lo05,wang05}.
Different intensities required for the decoy protocol are realised by intensity modulation, while the average intensity leaving Alice is set by the attenuator (Fig.~\ref{fig:setup}(b)).  In the decoy protocol, the photon fluxes are set as 0.5, 0.1 and 0.0007 photons per pulse with duty cycle of 98.8\%, 0.8\% and 0.4\% for signal and decoy states, respectively. The signal states are used for the generation of the secure keys whereas the weaker decoy pulses are used to protect the system from potential photon number splitting attacks. Information is encoded/decoded in the phase using the phase modulators.
The receiver is synchronized with the transmitter using the clock subsystem (Fig.~\ref{fig:setup}(a)). A feedback system is used to compensate both the drift in optical \textcolor{black}{polarization} and phase; this is accomplished through the use of a polarization controller and fiber stretcher respectively \cite{dixon10}. \textcolor{black}{The compensation operates continuously along with the key distribution, and there is no sacrifice in the duty cycle or the key rate.}

We determine the secure key rate from Koashi's security proof \cite{koashi06}, following the approach of Rice and Harrington \cite{rice09} to estimate single-photon parameters from decoy states. The secure key rate is given by
\begin{equation}\label{eq:secureR}
R=[Q_1(1-H(e_1))-Qf_{EC}(e)H(e)+Q_0]/t,
\end{equation}
where $Q_1$ is the estimated number of sifted bits from single-photon states, $e_1$ the estimated error rate of those states, $Q$ the total number of sifted bits, $f_{EC}$ the efficiency of the error correction, $e$ the QBER of sifted bits, $Q_0$ the estimated number of sifted arising from 0-photon pulses (dark and Raman noise counts), and $t$ the session duration. Each QKD session is sufficiently long for achieving a data block size greater than $5\times10^8$ bits. $H(x)=-\log_2x - (1-x)\log_2(1-x)$ is the binary entropy function. The single- and zero-photon quantities are estimated using a linear programming approach as given in Ref.~\cite{rice09}, with the difference that we neglect finite key effects and hence all upper and lower bounds are replaced with equalities.

\subsection{Clock subsystem.}
Accurate synchronization is important for QKD; especially high speed QKD with gated detectors. For synchronization in our QKD system, we use an off-the-shelf diode laser at Alice pulsed at 10~MHz, rather than the system clock rate of 1~GHz. The low pulsing rate allows a much lower clock laser launch power to be used, thus reducing the photon scatter into the quantum channel. A standard small-form-pluggable receiver at Bob detects the received clock before sending it to a frequency synthesizer to regenerate the original 1~GHz system clock at Bob.

\begin{figure}[t]
\newpage
\centering
\includegraphics[width=.8\columnwidth]{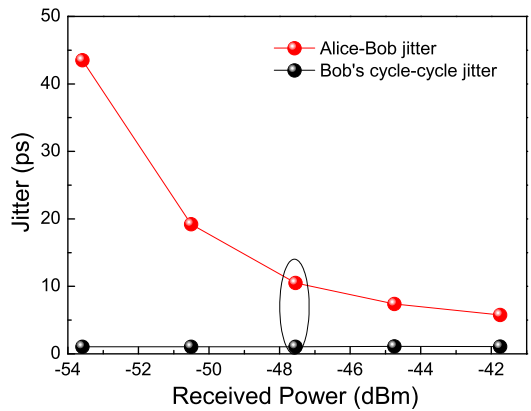}
\caption{Timing jitter \textit{vs.} received clock power measured between Alice and Bob over a fiber link of 80~km. Also shown is Bob's cycle-cycle jitter.}
\label{fig:jitter}
\newpage
\end{figure}

Figure~\ref{fig:jitter} shows the resulting timing jitter between Alice and Bob as a function of Bob's received clock power over 80~km of fiber. As the received power is increased, timing jitter gradually decreases.
However, there is a trade-off in reduced timing jitter and QKD system performance.
Excessive clock laser intensity results in increased photon scattering into the quantum channel raising the QBER. On the other hand, insufficient clock laser intensity reduces the effective detection efficiency of the quantum signal due to increased jitter between arriving laser pulses and the detector.
We decide to operate the clock subsystem at -47.6~dBm. This optical power is approximately 30 times smaller than either data-laser. The clock laser has thus negligible impact on the quantum channel.

The timing jitter between Alice and Bob is measured to be approximately 10~ps, a value sufficiently small to drive self-differencing detectors for efficient single-photon detection.  Note that these 10~ps include also the contribution from drift in the fiber, and thus represent the worst case for the recovered clock.
Measurement at Bob's side produces a cycle-cycle jitter of 1~ps throughout the power range used. This is indeed very low and ideal for driving detectors with the self-differencing technique.

\subsection{Raman-scattered light intensity}

Given a data-laser with optical power $I$ at a wavelength $\lambda_d$, its Raman-scattered light entering into the quantum receiver can be obtained by integrating over the entire fiber length ($L$) \cite{auyeung78,subacius05}, resulting in
\begin{equation}\label{eq:forward}
I_{Raman}^f=\beta(\lambda_d,\lambda_q,\delta)Ie^{-\alpha_qL}\int_0^Le^{(\alpha_q-\alpha_d)\ell}d\ell,
\end{equation}
\noindent and
\begin{equation}\label{eq:backward}
I_{Raman}^b=\beta(\lambda_d,\lambda_q,\delta)I(1-e^{-(\alpha_d+\alpha_q)L}),
\end{equation}
\noindent for the forward and backward reconfiguration, respectively. Here, $\lambda_q$ and $\delta$ are the central wavelength and bandwidth of the quantum channel, respectively, $\beta(\lambda_d,\lambda_q, \Delta)$ the Raman scatter coefficient, and $\alpha_d$ ($\alpha_q$) the fiber attenuation coefficient at a wavelength of $\lambda_d$($\lambda_q$).

The Raman scatter coefficient $\beta$ is measurable by the back-scattered Raman spectrum, see Fig.~\ref{fig:raman}(a) as an example.  We calculate the fiber length dependence of the Raman power, as shown by the solid lines in Fig.~\ref{fig:raman}(b).  The calculations agree well with the actual measurements.

\subsection{Simulation of the  secure key rate}

In order to simulate the secure key rate using Eqn.~\ref{eq:secureR}, we need to calculate the otherwise directly measurable quantities related to different classes of pulses used in the decoy protocol. These parameters include the QBER and transmittances. Transmittance is Bob's detection probability of a given class of pulses transmitted by Alice.

The QBER for the signal pulses ($e$) is approximated using
\begin{equation}\label{QBER}
e\doteq(e_{opt}+\frac{1}{2}P_a)+e_n,
\end{equation}
\noindent where $e_{opt}$ is due to encoding apparatus imperfections, such as finite interferometer visibility, misalignment and imperfect modulation, $P_a$ is the detector afterpulse probability, and $e_n$ is the noise contribution from both dark counts and Raman photons.  Both $e_{opt}$ and $P_a$ are fiber length independent, and give a combined contribution of 2.8\% to the QBER.

The fiber length dependent component of the QBER is written as
\begin{equation}\label{eq:e_n}
e_n=\frac{1}{2}\cdot \frac{P_d + P_R(L)}{\mu e^{-\alpha_q L}\eta_{Bob}+P_d+P_R(L)},
\end{equation}
\noindent where $P_d$ is the detector dark count probability, $P_R(L)$ the probability of registering a Raman photon per clock cycle, $\mu$ the photon flux of signal states, and $\eta_{Bob}$ Bob's detection efficiency. At each fiber length,  $P_R(L)$ is calculated using Eqns.~\ref{eq:forward} and \ref{eq:backward} with corrections from the data-laser power control, spectral and temporal filtering, and Bob's detection efficiency.

Excluding detector afterpulsing,  Bob's overall detection probability can be written as
\begin{equation}\label{eq:count}
T=\sum_{i=1}^3 P_i \mu_i e^{-\alpha_q L} \eta_{Bob} + [P_d + P_R(L)],
\end{equation}
\noindent where $P_i$ ($\sum_{i} P_i = 1$) is the probability that Alice transmits pulses with intensity of $\mu_i$. By including detector afterpulsing, we obtain the transmittance for each class of pulses:
\begin{equation}\label{eq:transmittance}
T_i= \mu_i e^{-\alpha_q L} \eta_{Bob} + [P_d + P_R(L)] + T \cdot P_a.
\end{equation}

We use $f_{EC}$=1.1 in the simulation.

\acknowledgments
The authors thank B. Fr\"ohlich for a critical reading of the manuscript and useful suggestions. K. A. Patel acknowledges personal support via the EPSRC funded CDT in Photonics System Development.



\begin{thebibliography}{34}%
\makeatletter
\providecommand \@ifxundefined [1]{%
 \@ifx{#1\undefined}
}%
\providecommand \@ifnum [1]{%
 \ifnum #1\expandafter \@firstoftwo
 \else \expandafter \@secondoftwo
 \fi
}%
\providecommand \@ifx [1]{%
 \ifx #1\expandafter \@firstoftwo
 \else \expandafter \@secondoftwo
 \fi
}%
\providecommand \natexlab [1]{#1}%
\providecommand \enquote  [1]{``#1''}%
\providecommand \bibnamefont  [1]{#1}%
\providecommand \bibfnamefont [1]{#1}%
\providecommand \citenamefont [1]{#1}%
\providecommand \href@noop [0]{\@secondoftwo}%
\providecommand \href [0]{\begingroup \@sanitize@url \@href}%
\providecommand \@href[1]{\@@startlink{#1}\@@href}%
\providecommand \@@href[1]{\endgroup#1\@@endlink}%
\providecommand \@sanitize@url [0]{\catcode `\\12\catcode `\$12\catcode
  `\&12\catcode `\#12\catcode `\^12\catcode `\_12\catcode `\%12\relax}%
\providecommand \@@startlink[1]{}%
\providecommand \@@endlink[0]{}%
\providecommand \url  [0]{\begingroup\@sanitize@url \@url }%
\providecommand \@url [1]{\endgroup\@href {#1}{\urlprefix }}%
\providecommand \urlprefix  [0]{URL }%
\providecommand \Eprint [0]{\href }%
\providecommand \doibase [0]{http://dx.doi.org/}%
\providecommand \selectlanguage [0]{\@gobble}%
\providecommand \bibinfo  [0]{\@secondoftwo}%
\providecommand \bibfield  [0]{\@secondoftwo}%
\providecommand \translation [1]{[#1]}%
\providecommand \BibitemOpen [0]{}%
\providecommand \bibitemStop [0]{}%
\providecommand \bibitemNoStop [0]{.\EOS\space}%
\providecommand \EOS [0]{\spacefactor3000\relax}%
\providecommand \BibitemShut  [1]{\csname bibitem#1\endcsname}%
\let\auto@bib@innerbib\@empty
\bibitem [{\citenamefont {Bennett}\ and\ \citenamefont
  {Brassard}(1984)}]{bb84}%
  \BibitemOpen
  \bibfield  {author} {\bibinfo {author} {\bibfnamefont {C.~H.}\ \bibnamefont
  {Bennett}}\ and\ \bibinfo {author} {\bibfnamefont {G.}~\bibnamefont
  {Brassard}},\ }\bibfield  {title} {\enquote {\bibinfo {title} {Quantum
  cryptography: public key distribution and coin tossing},}\ }in\ \href@noop {}
  {\emph {\bibinfo {booktitle} {Proceedings of the IEEE International
  Conference on Computers, Systems and Signal Processing}}}\ (\bibinfo
  {address} {Bangalore, India},\ \bibinfo {year} {1984})\ pp.\ \bibinfo {pages}
  {175--179}\BibitemShut {NoStop}%
\bibitem [{\citenamefont {Ekert}(1991)}]{ekert91}%
  \BibitemOpen
  \bibfield  {author} {\bibinfo {author} {\bibfnamefont {Artur~K.}\
  \bibnamefont {Ekert}},\ }\bibfield  {title} {\enquote {\bibinfo {title}
  {{Quantum cryptography based on Bell's theorem}},}\ }\href@noop {} {\bibfield
   {journal} {\bibinfo  {journal} {Phys. Rev. Lett.}\ }\textbf {\bibinfo
  {volume} {67}},\ \bibinfo {pages} {661--663} (\bibinfo {year}
  {1991})}\BibitemShut {NoStop}%
\bibitem [{\citenamefont {Elliot}\ \emph {et~al.}(2005)\citenamefont {Elliot},
  \citenamefont {Colvin}, \citenamefont {Pearson}, \citenamefont {Pikalo},
  \citenamefont {Schlafer},\ and\ \citenamefont {Yeh}}]{elliot05}%
  \BibitemOpen
  \bibfield  {author} {\bibinfo {author} {\bibfnamefont {C.}~\bibnamefont
  {Elliot}}, \bibinfo {author} {\bibfnamefont {A.}~\bibnamefont {Colvin}},
  \bibinfo {author} {\bibfnamefont {D.}~\bibnamefont {Pearson}}, \bibinfo
  {author} {\bibfnamefont {O.}~\bibnamefont {Pikalo}}, \bibinfo {author}
  {\bibfnamefont {J.}~\bibnamefont {Schlafer}}, \ and\ \bibinfo {author}
  {\bibfnamefont {H.}~\bibnamefont {Yeh}},\ }\href@noop {} {\enquote {\bibinfo
  {title} {Current status of the \uppercase{DAPRA} quantum network},}\ }
  (\bibinfo {year} {2005}),\ \bibinfo {note} {preprint:
  quant-ph/0503058}\BibitemShut {NoStop}%
\bibitem [{\citenamefont {Peev}\ \emph {et~al.}(2009)\citenamefont {Peev},
  \citenamefont {Pacher}, \citenamefont {All\'{e}aume}, \citenamefont
  {Barreiro}, \citenamefont {Bouda}, \citenamefont {Boxleitner}, \citenamefont
  {Debuisschert}, \citenamefont {Diamanti}, \citenamefont {Dianati},
  \citenamefont {Dynes}, \citenamefont {Fasel}, \citenamefont {Fossier},
  \citenamefont {F\"{u}rst}, \citenamefont {Gautier}, \citenamefont {Gay},
  \citenamefont {Gisin}, \citenamefont {Grangier}, \citenamefont {Happe},
  \citenamefont {Hasani}, \citenamefont {Hentschel}, \citenamefont {H\"{u}bel},
  \citenamefont {Humer}, \citenamefont {L\"{a}nger}, \citenamefont {Legr\'{e}},
  \citenamefont {Lieger}, \citenamefont {Lodewyck}, \citenamefont
  {Lor\"{u}nser}, \citenamefont {L\"{u}tkenhaus}, \citenamefont {Marhold},
  \citenamefont {Matyus}, \citenamefont {Maurhart}, \citenamefont {Monat},
  \citenamefont {Nauerth}, \citenamefont {Page}, \citenamefont {Poppe},
  \citenamefont {Querasser}, \citenamefont {Ribordy}, \citenamefont {Robyr},
  \citenamefont {Salvail}, \citenamefont {Sharpe}, \citenamefont {Shields},
  \citenamefont {Stucki}, \citenamefont {Suda}, \citenamefont {Tamas},
  \citenamefont {Themel}, \citenamefont {Thew}, \citenamefont {Thoma},
  \citenamefont {Treiber}, \citenamefont {Trinkler}, \citenamefont
  {Tualle-Brouri}, \citenamefont {Vannel}, \citenamefont {Walenta},
  \citenamefont {Weier}, \citenamefont {Weinfurter}, \citenamefont {Wimberger},
  \citenamefont {Yuan}, \citenamefont {Zbinden},\ and\ \citenamefont
  {Zeilinger}}]{peev09}%
  \BibitemOpen
  \bibfield  {author} {\bibinfo {author} {\bibfnamefont {M.}~\bibnamefont
  {Peev}}, \bibinfo {author} {\bibfnamefont {C.}~\bibnamefont {Pacher}},
  \bibinfo {author} {\bibfnamefont {R.}~\bibnamefont {All\'{e}aume}}, \bibinfo
  {author} {\bibfnamefont {C.}~\bibnamefont {Barreiro}}, \bibinfo {author}
  {\bibfnamefont {J.}~\bibnamefont {Bouda}}, \bibinfo {author} {\bibfnamefont
  {W.}~\bibnamefont {Boxleitner}}, \bibinfo {author} {\bibfnamefont
  {T.}~\bibnamefont {Debuisschert}}, \bibinfo {author} {\bibfnamefont
  {E.}~\bibnamefont {Diamanti}}, \bibinfo {author} {\bibfnamefont
  {M.}~\bibnamefont {Dianati}}, \bibinfo {author} {\bibfnamefont {J.~F.}\
  \bibnamefont {Dynes}}, \bibinfo {author} {\bibfnamefont {S.}~\bibnamefont
  {Fasel}}, \bibinfo {author} {\bibfnamefont {S.}~\bibnamefont {Fossier}},
  \bibinfo {author} {\bibfnamefont {M.}~\bibnamefont {F\"{u}rst}}, \bibinfo
  {author} {\bibfnamefont {J.~D.}\ \bibnamefont {Gautier}}, \bibinfo {author}
  {\bibfnamefont {O.}~\bibnamefont {Gay}}, \bibinfo {author} {\bibfnamefont
  {N.}~\bibnamefont {Gisin}}, \bibinfo {author} {\bibfnamefont
  {P.}~\bibnamefont {Grangier}}, \bibinfo {author} {\bibfnamefont
  {A.}~\bibnamefont {Happe}}, \bibinfo {author} {\bibfnamefont
  {Y.}~\bibnamefont {Hasani}}, \bibinfo {author} {\bibfnamefont
  {M.}~\bibnamefont {Hentschel}}, \bibinfo {author} {\bibfnamefont
  {H.}~\bibnamefont {H\"{u}bel}}, \bibinfo {author} {\bibfnamefont
  {G.}~\bibnamefont {Humer}}, \bibinfo {author} {\bibfnamefont
  {T.}~\bibnamefont {L\"{a}nger}}, \bibinfo {author} {\bibfnamefont
  {M.}~\bibnamefont {Legr\'{e}}}, \bibinfo {author} {\bibfnamefont
  {R.}~\bibnamefont {Lieger}}, \bibinfo {author} {\bibfnamefont
  {J.}~\bibnamefont {Lodewyck}}, \bibinfo {author} {\bibfnamefont
  {T.}~\bibnamefont {Lor\"{u}nser}}, \bibinfo {author} {\bibfnamefont
  {N.}~\bibnamefont {L\"{u}tkenhaus}}, \bibinfo {author} {\bibfnamefont
  {A.}~\bibnamefont {Marhold}}, \bibinfo {author} {\bibfnamefont
  {T.}~\bibnamefont {Matyus}}, \bibinfo {author} {\bibfnamefont
  {O.}~\bibnamefont {Maurhart}}, \bibinfo {author} {\bibfnamefont
  {L.}~\bibnamefont {Monat}}, \bibinfo {author} {\bibfnamefont
  {S.}~\bibnamefont {Nauerth}}, \bibinfo {author} {\bibfnamefont {J.~B.}\
  \bibnamefont {Page}}, \bibinfo {author} {\bibfnamefont {A.}~\bibnamefont
  {Poppe}}, \bibinfo {author} {\bibfnamefont {E.}~\bibnamefont {Querasser}},
  \bibinfo {author} {\bibfnamefont {G.}~\bibnamefont {Ribordy}}, \bibinfo
  {author} {\bibfnamefont {S.}~\bibnamefont {Robyr}}, \bibinfo {author}
  {\bibfnamefont {L.}~\bibnamefont {Salvail}}, \bibinfo {author} {\bibfnamefont
  {A.~W.}\ \bibnamefont {Sharpe}}, \bibinfo {author} {\bibfnamefont {A.~J.}\
  \bibnamefont {Shields}}, \bibinfo {author} {\bibfnamefont {D.}~\bibnamefont
  {Stucki}}, \bibinfo {author} {\bibfnamefont {M.}~\bibnamefont {Suda}},
  \bibinfo {author} {\bibfnamefont {C.}~\bibnamefont {Tamas}}, \bibinfo
  {author} {\bibfnamefont {T.}~\bibnamefont {Themel}}, \bibinfo {author}
  {\bibfnamefont {R.~T.}\ \bibnamefont {Thew}}, \bibinfo {author}
  {\bibfnamefont {Y.}~\bibnamefont {Thoma}}, \bibinfo {author} {\bibfnamefont
  {A.}~\bibnamefont {Treiber}}, \bibinfo {author} {\bibfnamefont
  {P.}~\bibnamefont {Trinkler}}, \bibinfo {author} {\bibfnamefont
  {R.}~\bibnamefont {Tualle-Brouri}}, \bibinfo {author} {\bibfnamefont
  {F.}~\bibnamefont {Vannel}}, \bibinfo {author} {\bibfnamefont
  {N.}~\bibnamefont {Walenta}}, \bibinfo {author} {\bibfnamefont
  {H.}~\bibnamefont {Weier}}, \bibinfo {author} {\bibfnamefont
  {H.}~\bibnamefont {Weinfurter}}, \bibinfo {author} {\bibfnamefont
  {I.}~\bibnamefont {Wimberger}}, \bibinfo {author} {\bibfnamefont {Z.~L.}\
  \bibnamefont {Yuan}}, \bibinfo {author} {\bibfnamefont {H.}~\bibnamefont
  {Zbinden}}, \ and\ \bibinfo {author} {\bibfnamefont {A.}~\bibnamefont
  {Zeilinger}},\ }\bibfield  {title} {\enquote {\bibinfo {title} {The
  \uppercase{SECOQC} quantum key distribution network in \uppercase{V}ienna},}\
  }\href@noop {} {\bibfield  {journal} {\bibinfo  {journal} {New J. Phys.}\
  }\textbf {\bibinfo {volume} {11}},\ \bibinfo {eid} {075001} (\bibinfo {year}
  {2009})}\BibitemShut {NoStop}%
\bibitem [{\citenamefont {Sasaki}\ \emph {et~al.}(2011)\citenamefont {Sasaki},
  \citenamefont {Fujiwara}, \citenamefont {Ishizuka}, \citenamefont {Klaus},
  \citenamefont {Wakui}, \citenamefont {Takeoka}, \citenamefont {Miki},
  \citenamefont {Yamashita}, \citenamefont {Wang}, \citenamefont {Tanaka},
  \citenamefont {Yoshino}, \citenamefont {Nambu}, \citenamefont {Takahashi},
  \citenamefont {Tajima}, \citenamefont {Tomita}, \citenamefont {Domeki},
  \citenamefont {Hasegawa}, \citenamefont {Sakai}, \citenamefont {Kobayashi},
  \citenamefont {Asai}, \citenamefont {Shimizu}, \citenamefont {Tokura},
  \citenamefont {Tsurumaru}, \citenamefont {Matsui}, \citenamefont {Honjo},
  \citenamefont {Tamaki}, \citenamefont {Takesue}, \citenamefont {Tokura},
  \citenamefont {Dynes}, \citenamefont {Dixon}, \citenamefont {Sharpe},
  \citenamefont {Yuan}, \citenamefont {Shields}, \citenamefont {Uchikoga},
  \citenamefont {Legr\'{e}}, \citenamefont {Robyr}, \citenamefont {Trinkler},
  \citenamefont {Monat}, \citenamefont {Page}, \citenamefont {Ribordy},
  \citenamefont {Poppe}, \citenamefont {Allacher}, \citenamefont {Maurhart},
  \citenamefont {L\"{a}nger}, \citenamefont {Peev},\ and\ \citenamefont
  {Zeilinger}}]{sasaki11}%
  \BibitemOpen
  \bibfield  {author} {\bibinfo {author} {\bibfnamefont {M.}~\bibnamefont
  {Sasaki}}, \bibinfo {author} {\bibfnamefont {M.}~\bibnamefont {Fujiwara}},
  \bibinfo {author} {\bibfnamefont {H.}~\bibnamefont {Ishizuka}}, \bibinfo
  {author} {\bibfnamefont {W.}~\bibnamefont {Klaus}}, \bibinfo {author}
  {\bibfnamefont {K.}~\bibnamefont {Wakui}}, \bibinfo {author} {\bibfnamefont
  {M.}~\bibnamefont {Takeoka}}, \bibinfo {author} {\bibfnamefont
  {S.}~\bibnamefont {Miki}}, \bibinfo {author} {\bibfnamefont {T.}~\bibnamefont
  {Yamashita}}, \bibinfo {author} {\bibfnamefont {Z.}~\bibnamefont {Wang}},
  \bibinfo {author} {\bibfnamefont {A.}~\bibnamefont {Tanaka}}, \bibinfo
  {author} {\bibfnamefont {K.}~\bibnamefont {Yoshino}}, \bibinfo {author}
  {\bibfnamefont {Y.}~\bibnamefont {Nambu}}, \bibinfo {author} {\bibfnamefont
  {S.}~\bibnamefont {Takahashi}}, \bibinfo {author} {\bibfnamefont
  {A.}~\bibnamefont {Tajima}}, \bibinfo {author} {\bibfnamefont
  {A.}~\bibnamefont {Tomita}}, \bibinfo {author} {\bibfnamefont
  {T.}~\bibnamefont {Domeki}}, \bibinfo {author} {\bibfnamefont
  {T.}~\bibnamefont {Hasegawa}}, \bibinfo {author} {\bibfnamefont
  {Y.}~\bibnamefont {Sakai}}, \bibinfo {author} {\bibfnamefont
  {H.}~\bibnamefont {Kobayashi}}, \bibinfo {author} {\bibfnamefont
  {T.}~\bibnamefont {Asai}}, \bibinfo {author} {\bibfnamefont {K.}~\bibnamefont
  {Shimizu}}, \bibinfo {author} {\bibfnamefont {T.}~\bibnamefont {Tokura}},
  \bibinfo {author} {\bibfnamefont {T.}~\bibnamefont {Tsurumaru}}, \bibinfo
  {author} {\bibfnamefont {M.}~\bibnamefont {Matsui}}, \bibinfo {author}
  {\bibfnamefont {T.}~\bibnamefont {Honjo}}, \bibinfo {author} {\bibfnamefont
  {K.}~\bibnamefont {Tamaki}}, \bibinfo {author} {\bibfnamefont
  {H.}~\bibnamefont {Takesue}}, \bibinfo {author} {\bibfnamefont
  {Y.}~\bibnamefont {Tokura}}, \bibinfo {author} {\bibfnamefont {J.~F.}\
  \bibnamefont {Dynes}}, \bibinfo {author} {\bibfnamefont {A.~R.}\ \bibnamefont
  {Dixon}}, \bibinfo {author} {\bibfnamefont {A.~W.}\ \bibnamefont {Sharpe}},
  \bibinfo {author} {\bibfnamefont {Z.~L.}\ \bibnamefont {Yuan}}, \bibinfo
  {author} {\bibfnamefont {A.~J.}\ \bibnamefont {Shields}}, \bibinfo {author}
  {\bibfnamefont {S.}~\bibnamefont {Uchikoga}}, \bibinfo {author}
  {\bibfnamefont {M.}~\bibnamefont {Legr\'{e}}}, \bibinfo {author}
  {\bibfnamefont {S.}~\bibnamefont {Robyr}}, \bibinfo {author} {\bibfnamefont
  {P.}~\bibnamefont {Trinkler}}, \bibinfo {author} {\bibfnamefont
  {L.}~\bibnamefont {Monat}}, \bibinfo {author} {\bibfnamefont {J.-B.}\
  \bibnamefont {Page}}, \bibinfo {author} {\bibfnamefont {G.}~\bibnamefont
  {Ribordy}}, \bibinfo {author} {\bibfnamefont {A.}~\bibnamefont {Poppe}},
  \bibinfo {author} {\bibfnamefont {A.}~\bibnamefont {Allacher}}, \bibinfo
  {author} {\bibfnamefont {O.}~\bibnamefont {Maurhart}}, \bibinfo {author}
  {\bibfnamefont {T.}~\bibnamefont {L\"{a}nger}}, \bibinfo {author}
  {\bibfnamefont {M.}~\bibnamefont {Peev}}, \ and\ \bibinfo {author}
  {\bibfnamefont {A.}~\bibnamefont {Zeilinger}},\ }\bibfield  {title} {\enquote
  {\bibinfo {title} {Field test of quantum key distribution in the
  \uppercase{T}okyo \uppercase{QKD} network},}\ }\href@noop {} {\bibfield
  {journal} {\bibinfo  {journal} {Opt. Express}\ }\textbf {\bibinfo {volume}
  {19}},\ \bibinfo {pages} {10387--10409} (\bibinfo {year} {2011})}\BibitemShut
  {NoStop}%
\bibitem [{\citenamefont {Dixon}\ \emph {et~al.}(2008)\citenamefont {Dixon},
  \citenamefont {Yuan}, \citenamefont {Dynes}, \citenamefont {Sharpe},\ and\
  \citenamefont {Shields}}]{dixon08}%
  \BibitemOpen
  \bibfield  {author} {\bibinfo {author} {\bibfnamefont {A.~R.}\ \bibnamefont
  {Dixon}}, \bibinfo {author} {\bibfnamefont {Z.~L.}\ \bibnamefont {Yuan}},
  \bibinfo {author} {\bibfnamefont {J.~F.}\ \bibnamefont {Dynes}}, \bibinfo
  {author} {\bibfnamefont {A.~W.}\ \bibnamefont {Sharpe}}, \ and\ \bibinfo
  {author} {\bibfnamefont {A.~J.}\ \bibnamefont {Shields}},\ }\bibfield
  {title} {\enquote {\bibinfo {title} {Gigahertz decoy quantum key distribution
  with 1 \uppercase{M}bit/s secure key rate},}\ }\href@noop {} {\bibfield
  {journal} {\bibinfo  {journal} {Optics Express}\ }\textbf {\bibinfo {volume}
  {16}},\ \bibinfo {pages} {18790--18979} (\bibinfo {year} {2008})}\BibitemShut
  {NoStop}%
\bibitem [{\citenamefont {Zhang}\ \emph {et~al.}({2009})\citenamefont {Zhang},
  \citenamefont {Takesue}, \citenamefont {Honjo}, \citenamefont {Wen},
  \citenamefont {Hirohata}, \citenamefont {Suyama}, \citenamefont {Takiguchi},
  \citenamefont {Kamada}, \citenamefont {Tokura}, \citenamefont {Tadanaga},
  \citenamefont {Nishida}, \citenamefont {Asobe},\ and\ \citenamefont
  {Yamamoto}}]{zhang09}%
  \BibitemOpen
  \bibfield  {author} {\bibinfo {author} {\bibfnamefont {Q.}~\bibnamefont
  {Zhang}}, \bibinfo {author} {\bibfnamefont {H.}~\bibnamefont {Takesue}},
  \bibinfo {author} {\bibfnamefont {T.}~\bibnamefont {Honjo}}, \bibinfo
  {author} {\bibfnamefont {K.}~\bibnamefont {Wen}}, \bibinfo {author}
  {\bibfnamefont {T.}~\bibnamefont {Hirohata}}, \bibinfo {author}
  {\bibfnamefont {M.}~\bibnamefont {Suyama}}, \bibinfo {author} {\bibfnamefont
  {Y.}~\bibnamefont {Takiguchi}}, \bibinfo {author} {\bibfnamefont
  {H.}~\bibnamefont {Kamada}}, \bibinfo {author} {\bibfnamefont
  {Y.}~\bibnamefont {Tokura}}, \bibinfo {author} {\bibfnamefont
  {O.}~\bibnamefont {Tadanaga}}, \bibinfo {author} {\bibfnamefont
  {Y.}~\bibnamefont {Nishida}}, \bibinfo {author} {\bibfnamefont
  {M.}~\bibnamefont {Asobe}}, \ and\ \bibinfo {author} {\bibfnamefont
  {Y.}~\bibnamefont {Yamamoto}},\ }\bibfield  {title} {\enquote {\bibinfo
  {title} {{Megabits secure key rate quantum key distribution}},}\ }\href@noop
  {} {\bibfield  {journal} {\bibinfo  {journal} {{New. J. Phys.}}\ }\textbf
  {\bibinfo {volume} {{11}}},\ \bibinfo {pages} {{045010}} (\bibinfo {year}
  {{2009}})}\BibitemShut {NoStop}%
\bibitem [{\citenamefont {Dixon}\ \emph {et~al.}(2010)\citenamefont {Dixon},
  \citenamefont {Yuan}, \citenamefont {Dynes}, \citenamefont {Sharpe},\ and\
  \citenamefont {Shields}}]{dixon10}%
  \BibitemOpen
  \bibfield  {author} {\bibinfo {author} {\bibfnamefont {A.~R.}\ \bibnamefont
  {Dixon}}, \bibinfo {author} {\bibfnamefont {Z.~L.}\ \bibnamefont {Yuan}},
  \bibinfo {author} {\bibfnamefont {J.~F.}\ \bibnamefont {Dynes}}, \bibinfo
  {author} {\bibfnamefont {A.~W.}\ \bibnamefont {Sharpe}}, \ and\ \bibinfo
  {author} {\bibfnamefont {A.~J.}\ \bibnamefont {Shields}},\ }\bibfield
  {title} {\enquote {\bibinfo {title} {Continuous operation of high bit rate
  quantum key distribution},}\ }\href@noop {} {\bibfield  {journal} {\bibinfo
  {journal} {Appl. Phys. Lett.}\ }\textbf {\bibinfo {volume} {96}},\ \bibinfo
  {eid} {161102} (\bibinfo {year} {2010})}\BibitemShut {NoStop}%
\bibitem [{\citenamefont {Stucki}\ \emph {et~al.}(2009)\citenamefont {Stucki},
  \citenamefont {Walenta}, \citenamefont {Vannel}, \citenamefont {Thew},
  \citenamefont {Gisin}, \citenamefont {Zbinden}, \citenamefont {Gray},
  \citenamefont {Towery},\ and\ \citenamefont {Ten}}]{stucki09b}%
  \BibitemOpen
  \bibfield  {author} {\bibinfo {author} {\bibfnamefont {D.}~\bibnamefont
  {Stucki}}, \bibinfo {author} {\bibfnamefont {N.}~\bibnamefont {Walenta}},
  \bibinfo {author} {\bibfnamefont {F.}~\bibnamefont {Vannel}}, \bibinfo
  {author} {\bibfnamefont {R.~T.}\ \bibnamefont {Thew}}, \bibinfo {author}
  {\bibfnamefont {N.}~\bibnamefont {Gisin}}, \bibinfo {author} {\bibfnamefont
  {H.}~\bibnamefont {Zbinden}}, \bibinfo {author} {\bibfnamefont
  {S.}~\bibnamefont {Gray}}, \bibinfo {author} {\bibfnamefont {C.~R.}\
  \bibnamefont {Towery}}, \ and\ \bibinfo {author} {\bibfnamefont
  {S.}~\bibnamefont {Ten}},\ }\bibfield  {title} {\enquote {\bibinfo {title}
  {High rate, long-distance quantum key distribution over 250 km of ultra low
  loss fibres},}\ }\href@noop {} {\bibfield  {journal} {\bibinfo  {journal}
  {New J. Phys.}\ }\textbf {\bibinfo {volume} {11}},\ \bibinfo {pages} {075003}
  (\bibinfo {year} {2009})}\BibitemShut {NoStop}%
\bibitem [{\citenamefont {Wang}\ \emph {et~al.}(2012)\citenamefont {Wang},
  \citenamefont {Chen}, \citenamefont {Guo}, \citenamefont {Yin}, \citenamefont
  {Li}, \citenamefont {Zhou}, \citenamefont {Guo},\ and\ \citenamefont
  {Han}}]{wang12}%
  \BibitemOpen
  \bibfield  {author} {\bibinfo {author} {\bibfnamefont {S.}~\bibnamefont
  {Wang}}, \bibinfo {author} {\bibfnamefont {W.}~\bibnamefont {Chen}}, \bibinfo
  {author} {\bibfnamefont {J.~F.}\ \bibnamefont {Guo}}, \bibinfo {author}
  {\bibfnamefont {Z.~Q.}\ \bibnamefont {Yin}}, \bibinfo {author} {\bibfnamefont
  {H.~W.}\ \bibnamefont {Li}}, \bibinfo {author} {\bibfnamefont
  {Z.}~\bibnamefont {Zhou}}, \bibinfo {author} {\bibfnamefont {G.~C.}\
  \bibnamefont {Guo}}, \ and\ \bibinfo {author} {\bibfnamefont {Z.~F.}\
  \bibnamefont {Han}},\ }\bibfield  {title} {\enquote {\bibinfo {title} {2
  \uppercase{GH}z clock quantum key distribution over 260~km of standard
  telecom fiber},}\ }\href@noop {} {\bibfield  {journal} {\bibinfo  {journal}
  {Opt. Lett.}\ }\textbf {\bibinfo {volume} {37}},\ \bibinfo {pages}
  {1008--1010} (\bibinfo {year} {2012})}\BibitemShut {NoStop}%
\bibitem [{\citenamefont {Townsend}(1997)}]{townsend97}%
  \BibitemOpen
  \bibfield  {author} {\bibinfo {author} {\bibfnamefont {P.~D.}\ \bibnamefont
  {Townsend}},\ }\bibfield  {title} {\enquote {\bibinfo {title} {Simultaneous
  quantum cryptographic key distribution and conventional data transmission
  over installed fibre using wavelength-division multiplexing},}\ }\href@noop
  {} {\bibfield  {journal} {\bibinfo  {journal} {Electron. Lett.}\ }\textbf
  {\bibinfo {volume} {33}},\ \bibinfo {pages} {188--190} (\bibinfo {year}
  {1997})}\BibitemShut {NoStop}%
\bibitem [{\citenamefont {Chapuran}\ \emph {et~al.}(2009)\citenamefont
  {Chapuran}, \citenamefont {Toliver}, \citenamefont {Peters}, \citenamefont
  {Jackel}, \citenamefont {Goodman}, \citenamefont {Runser}, \citenamefont
  {McNown}, \citenamefont {Dallmann}, \citenamefont {Hughes}, \citenamefont
  {McCabe}, \citenamefont {Nordholt}, \citenamefont {Peterson}, \citenamefont
  {Tyagi}, \citenamefont {Mercer},\ and\ \citenamefont {Dardy}}]{chapuran09}%
  \BibitemOpen
  \bibfield  {author} {\bibinfo {author} {\bibfnamefont {T.~E.}\ \bibnamefont
  {Chapuran}}, \bibinfo {author} {\bibfnamefont {P.}~\bibnamefont {Toliver}},
  \bibinfo {author} {\bibfnamefont {N.~A.}\ \bibnamefont {Peters}}, \bibinfo
  {author} {\bibfnamefont {J.}~\bibnamefont {Jackel}}, \bibinfo {author}
  {\bibfnamefont {M.~S.}\ \bibnamefont {Goodman}}, \bibinfo {author}
  {\bibfnamefont {R.~J.}\ \bibnamefont {Runser}}, \bibinfo {author}
  {\bibfnamefont {S.~R.}\ \bibnamefont {McNown}}, \bibinfo {author}
  {\bibfnamefont {N.}~\bibnamefont {Dallmann}}, \bibinfo {author}
  {\bibfnamefont {R.~J.}\ \bibnamefont {Hughes}}, \bibinfo {author}
  {\bibfnamefont {K.~P.}\ \bibnamefont {McCabe}}, \bibinfo {author}
  {\bibfnamefont {J.~E.}\ \bibnamefont {Nordholt}}, \bibinfo {author}
  {\bibfnamefont {C.~G.}\ \bibnamefont {Peterson}}, \bibinfo {author}
  {\bibfnamefont {K.~T.}\ \bibnamefont {Tyagi}}, \bibinfo {author}
  {\bibfnamefont {L.}~\bibnamefont {Mercer}}, \ and\ \bibinfo {author}
  {\bibfnamefont {H.}~\bibnamefont {Dardy}},\ }\bibfield  {title} {\enquote
  {\bibinfo {title} {Optical networking for quantum key distribution and
  quantum communications},}\ }\href@noop {} {\bibfield  {journal} {\bibinfo
  {journal} {New J. Phys.}\ }\textbf {\bibinfo {volume} {11}},\ \bibinfo
  {pages} {105001} (\bibinfo {year} {2009})}\BibitemShut {NoStop}%
\bibitem [{\citenamefont {Choi}\ \emph {et~al.}(2011)\citenamefont {Choi},
  \citenamefont {Young},\ and\ \citenamefont {Townsend}}]{choi11}%
  \BibitemOpen
  \bibfield  {author} {\bibinfo {author} {\bibfnamefont {I.}~\bibnamefont
  {Choi}}, \bibinfo {author} {\bibfnamefont {R.~J.}\ \bibnamefont {Young}}, \
  and\ \bibinfo {author} {\bibfnamefont {P.~D.}\ \bibnamefont {Townsend}},\
  }\bibfield  {title} {\enquote {\bibinfo {title} {Quantum information to the
  home},}\ }\href@noop {} {\bibfield  {journal} {\bibinfo  {journal} {New. J.
  Phys.}\ }\textbf {\bibinfo {volume} {13}},\ \bibinfo {eid} {063039} (\bibinfo
  {year} {2011})}\BibitemShut {NoStop}%
\bibitem [{\citenamefont {Eraerds}\ \emph {et~al.}(2010)\citenamefont
  {Eraerds}, \citenamefont {Walenta}, \citenamefont {Legr\'{e}}, \citenamefont
  {Gisin},\ and\ \citenamefont {Zbinden}}]{eraerds10}%
  \BibitemOpen
  \bibfield  {author} {\bibinfo {author} {\bibfnamefont {P.}~\bibnamefont
  {Eraerds}}, \bibinfo {author} {\bibfnamefont {N.}~\bibnamefont {Walenta}},
  \bibinfo {author} {\bibfnamefont {M.}~\bibnamefont {Legr\'{e}}}, \bibinfo
  {author} {\bibfnamefont {N.}~\bibnamefont {Gisin}}, \ and\ \bibinfo {author}
  {\bibfnamefont {H.}~\bibnamefont {Zbinden}},\ }\bibfield  {title} {\enquote
  {\bibinfo {title} {Quantum key distribution and 1 \uppercase{G}bps data
  encryption over a single fibre},}\ }\href@noop {} {\bibfield  {journal}
  {\bibinfo  {journal} {New J. Phys.}\ }\textbf {\bibinfo {volume} {12}},\
  \bibinfo {pages} {063027} (\bibinfo {year} {2010})}\BibitemShut {NoStop}%
\bibitem [{\citenamefont {Lancho}\ \emph {et~al.}(2010)\citenamefont {Lancho},
  \citenamefont {Martinez}, \citenamefont {Elkouss}, \citenamefont {Soto},\
  and\ \citenamefont {Martin}}]{lancho10}%
  \BibitemOpen
  \bibfield  {author} {\bibinfo {author} {\bibfnamefont {D.}~\bibnamefont
  {Lancho}}, \bibinfo {author} {\bibfnamefont {J.}~\bibnamefont {Martinez}},
  \bibinfo {author} {\bibfnamefont {D.}~\bibnamefont {Elkouss}}, \bibinfo
  {author} {\bibfnamefont {M.}~\bibnamefont {Soto}}, \ and\ \bibinfo {author}
  {\bibfnamefont {V.}~\bibnamefont {Martin}},\ }\bibfield  {title} {\enquote
  {\bibinfo {title} {\uppercase{QKD} in standard optical telecommunication
  networks},}\ }\href@noop {} {\bibfield  {journal} {\bibinfo  {journal} {Lect.
  Notes Inst. Comput. Sci., Soc. Inf. Telecom. Eng.}\ }\textbf {\bibinfo
  {volume} {36}},\ \bibinfo {pages} {142--149} (\bibinfo {year}
  {2010})}\BibitemShut {NoStop}%
\bibitem [{\citenamefont {Qi}\ \emph {et~al.}(2010)\citenamefont {Qi},
  \citenamefont {Zhu}, \citenamefont {Qian},\ and\ \citenamefont {Lo}}]{qi10}%
  \BibitemOpen
  \bibfield  {author} {\bibinfo {author} {\bibfnamefont {B.}~\bibnamefont
  {Qi}}, \bibinfo {author} {\bibfnamefont {W.}~\bibnamefont {Zhu}}, \bibinfo
  {author} {\bibfnamefont {L.}~\bibnamefont {Qian}}, \ and\ \bibinfo {author}
  {\bibfnamefont {H.~K.}\ \bibnamefont {Lo}},\ }\bibfield  {title} {\enquote
  {\bibinfo {title} {Feasibility of quantum key distribution through a dense
  wavelength division multiplexing network},}\ }\href@noop {} {\bibfield
  {journal} {\bibinfo  {journal} {New J. Phys.}\ }\textbf {\bibinfo {volume}
  {12}},\ \bibinfo {pages} {103042} (\bibinfo {year} {2010})}\BibitemShut
  {NoStop}%
\bibitem [{\citenamefont {Lo}\ \emph {et~al.}(2005)\citenamefont {Lo},
  \citenamefont {Ma},\ and\ \citenamefont {Chen}}]{lo05}%
  \BibitemOpen
  \bibfield  {author} {\bibinfo {author} {\bibfnamefont {H.~K.}\ \bibnamefont
  {Lo}}, \bibinfo {author} {\bibfnamefont {X.~F.}\ \bibnamefont {Ma}}, \ and\
  \bibinfo {author} {\bibfnamefont {K.}~\bibnamefont {Chen}},\ }\bibfield
  {title} {\enquote {\bibinfo {title} {Decoy state quantum key distribution},}\
  }\href@noop {} {\bibfield  {journal} {\bibinfo  {journal} {Phys. Rev. Lett.}\
  }\textbf {\bibinfo {volume} {94}},\ \bibinfo {pages} {230504} (\bibinfo
  {year} {2005})}\BibitemShut {NoStop}%
\bibitem [{\citenamefont {Wang}(2005)}]{wang05}%
  \BibitemOpen
  \bibfield  {author} {\bibinfo {author} {\bibfnamefont {X.~B.}\ \bibnamefont
  {Wang}},\ }\bibfield  {title} {\enquote {\bibinfo {title} {Beating the
  photon-number-splitting attack in practical quantum cryptography},}\
  }\href@noop {} {\bibfield  {journal} {\bibinfo  {journal} {Phys. Rev. Lett.}\
  }\textbf {\bibinfo {volume} {94}},\ \bibinfo {eid} {230503} (\bibinfo {year}
  {2005})}\BibitemShut {NoStop}%
\bibitem [{\citenamefont {Fernandez}\ \emph {et~al.}(2007)\citenamefont
  {Fernandez}, \citenamefont {Collins}, \citenamefont {Gordon}, \citenamefont
  {Townsend},\ and\ \citenamefont {Buller}}]{fernandez07}%
  \BibitemOpen
  \bibfield  {author} {\bibinfo {author} {\bibfnamefont {V.}~\bibnamefont
  {Fernandez}}, \bibinfo {author} {\bibfnamefont {R.~J.}\ \bibnamefont
  {Collins}}, \bibinfo {author} {\bibfnamefont {K.~J.}\ \bibnamefont {Gordon}},
  \bibinfo {author} {\bibfnamefont {P~D}\ \bibnamefont {Townsend}}, \ and\
  \bibinfo {author} {\bibfnamefont {G.~S.}\ \bibnamefont {Buller}},\ }\bibfield
   {title} {\enquote {\bibinfo {title} {Passive optical network approach to
  gigahertz-clocked multiuser quantum key distribution},}\ }\href@noop {}
  {\bibfield  {journal} {\bibinfo  {journal} {IEEE J. Quant. Electron.}\
  }\textbf {\bibinfo {volume} {43}},\ \bibinfo {pages} {130--138} (\bibinfo
  {year} {2007})}\BibitemShut {NoStop}%
\bibitem [{\citenamefont {Holloway}\ \emph {et~al.}(2011)\citenamefont
  {Holloway}, \citenamefont {Meyer-Scott}, \citenamefont {Erven},\ and\
  \citenamefont {Jennewein}}]{holloway11}%
  \BibitemOpen
  \bibfield  {author} {\bibinfo {author} {\bibfnamefont {C.}~\bibnamefont
  {Holloway}}, \bibinfo {author} {\bibfnamefont {E.}~\bibnamefont
  {Meyer-Scott}}, \bibinfo {author} {\bibfnamefont {C.}~\bibnamefont {Erven}},
  \ and\ \bibinfo {author} {\bibfnamefont {T.}~\bibnamefont {Jennewein}},\
  }\bibfield  {title} {\enquote {\bibinfo {title} {Quantum entanglement
  distribution with 810~nm photons through active telecommunication fibers},}\
  }\href@noop {} {\bibfield  {journal} {\bibinfo  {journal} {Opt. Express}\
  }\textbf {\bibinfo {volume} {19}},\ \bibinfo {pages} {20597--20603} (\bibinfo
  {year} {2011})}\BibitemShut {NoStop}%
\bibitem [{\citenamefont {Yuan}\ \emph {et~al.}(2007)\citenamefont {Yuan},
  \citenamefont {Kardynal}, \citenamefont {Sharpe},\ and\ \citenamefont
  {Shields}}]{yuan07}%
  \BibitemOpen
  \bibfield  {author} {\bibinfo {author} {\bibfnamefont {Z.~L.}\ \bibnamefont
  {Yuan}}, \bibinfo {author} {\bibfnamefont {B.~E.}\ \bibnamefont {Kardynal}},
  \bibinfo {author} {\bibfnamefont {A.~W.}\ \bibnamefont {Sharpe}}, \ and\
  \bibinfo {author} {\bibfnamefont {A.~J.}\ \bibnamefont {Shields}},\
  }\bibfield  {title} {\enquote {\bibinfo {title} {High speed single photon
  detection in the near infrared},}\ }\href@noop {} {\bibfield  {journal}
  {\bibinfo  {journal} {Appl. Phys. Lett.}\ }\textbf {\bibinfo {volume} {91}},\
  \bibinfo {eid} {041114} (\bibinfo {year} {2007})}\BibitemShut {NoStop}%
\bibitem [{\citenamefont {Yuan}\ \emph {et~al.}(2008)\citenamefont {Yuan},
  \citenamefont {Dixon}, \citenamefont {Dynes}, \citenamefont {Sharpe},\ and\
  \citenamefont {Shields}}]{yuan08}%
  \BibitemOpen
  \bibfield  {author} {\bibinfo {author} {\bibfnamefont {Z.~L.}\ \bibnamefont
  {Yuan}}, \bibinfo {author} {\bibfnamefont {A.~R.}\ \bibnamefont {Dixon}},
  \bibinfo {author} {\bibfnamefont {J.~F.}\ \bibnamefont {Dynes}}, \bibinfo
  {author} {\bibfnamefont {A.~W.}\ \bibnamefont {Sharpe}}, \ and\ \bibinfo
  {author} {\bibfnamefont {A.~J.}\ \bibnamefont {Shields}},\ }\bibfield
  {title} {\enquote {\bibinfo {title} {Gigahertz quantum key distribution with
  \uppercase{I}n\uppercase{G}a\uppercase{A}s avalanche photodiodes},}\
  }\href@noop {} {\bibfield  {journal} {\bibinfo  {journal} {Appl. Phys.
  Lett.}\ }\textbf {\bibinfo {volume} {92}},\ \bibinfo {eid} {201104} (\bibinfo
  {year} {2008})}\BibitemShut {NoStop}%
\bibitem [{\citenamefont {Dixon}\ \emph {et~al.}(2009)\citenamefont {Dixon},
  \citenamefont {Dynes}, \citenamefont {Yuan}, \citenamefont {Sharpe},
  \citenamefont {Bennett},\ and\ \citenamefont {Shields}}]{dixon09}%
  \BibitemOpen
  \bibfield  {author} {\bibinfo {author} {\bibfnamefont {A.~R.}\ \bibnamefont
  {Dixon}}, \bibinfo {author} {\bibfnamefont {J.~F.}\ \bibnamefont {Dynes}},
  \bibinfo {author} {\bibfnamefont {Z.~L.}\ \bibnamefont {Yuan}}, \bibinfo
  {author} {\bibfnamefont {A.~W.}\ \bibnamefont {Sharpe}}, \bibinfo {author}
  {\bibfnamefont {A.~J.}\ \bibnamefont {Bennett}}, \ and\ \bibinfo {author}
  {\bibfnamefont {A.~J.}\ \bibnamefont {Shields}},\ }\bibfield  {title}
  {\enquote {\bibinfo {title} {Ultrashort dead time of photon-counting
  \uppercase{I}n\uppercase{G}a\uppercase{A}s avalanche photodiodes},}\
  }\href@noop {} {\bibfield  {journal} {\bibinfo  {journal} {Appl. Phys.
  Lett.}\ }\textbf {\bibinfo {volume} {94}},\ \bibinfo {pages} {231113}
  (\bibinfo {year} {2009})}\BibitemShut {NoStop}%
\bibitem [{\citenamefont {Patel}\ \emph {et~al.}(2012)\citenamefont {Patel},
  \citenamefont {Dynes}, \citenamefont {Sharpe}, \citenamefont {Yuan},
  \citenamefont {Penty},\ and\ \citenamefont {Shields}}]{patel12}%
  \BibitemOpen
  \bibfield  {author} {\bibinfo {author} {\bibfnamefont {K.~A.}\ \bibnamefont
  {Patel}}, \bibinfo {author} {\bibfnamefont {J.~F.}\ \bibnamefont {Dynes}},
  \bibinfo {author} {\bibfnamefont {A.~W.}\ \bibnamefont {Sharpe}}, \bibinfo
  {author} {\bibfnamefont {Z.~L.}\ \bibnamefont {Yuan}}, \bibinfo {author}
  {\bibfnamefont {R.~V.}\ \bibnamefont {Penty}}, \ and\ \bibinfo {author}
  {\bibfnamefont {A.~J.}\ \bibnamefont {Shields}},\ }\bibfield  {title}
  {\enquote {\bibinfo {title} {Gigacount/second photon detection with
  \uppercase{I}n\uppercase{G}a\uppercase{A}s avalanche photodiodes},}\
  }\href@noop {} {\bibfield  {journal} {\bibinfo  {journal} {Electron. Lett.}\
  }\textbf {\bibinfo {volume} {48}},\ \bibinfo {pages} {111--113} (\bibinfo
  {year} {2012})}\BibitemShut {NoStop}%
\bibitem [{\citenamefont {Yuan}\ \emph {et~al.}(2010)\citenamefont {Yuan},
  \citenamefont {Dynes}, \citenamefont {Sharpe},\ and\ \citenamefont
  {Shields}}]{yuan10c}%
  \BibitemOpen
  \bibfield  {author} {\bibinfo {author} {\bibfnamefont {Z.~L.}\ \bibnamefont
  {Yuan}}, \bibinfo {author} {\bibfnamefont {J.~F.}\ \bibnamefont {Dynes}},
  \bibinfo {author} {\bibfnamefont {A.~W.}\ \bibnamefont {Sharpe}}, \ and\
  \bibinfo {author} {\bibfnamefont {A.~J.}\ \bibnamefont {Shields}},\
  }\bibfield  {title} {\enquote {\bibinfo {title} {Evolution of locally excited
  avalanches in semiconductors},}\ }\href@noop {} {\bibfield  {journal}
  {\bibinfo  {journal} {Appl. Phys. Lett.}\ }\textbf {\bibinfo {volume} {96}},\
  \bibinfo {pages} {191107} (\bibinfo {year} {2010})}\BibitemShut {NoStop}%
\bibitem [{\citenamefont {Koashi}(2006)}]{koashi06}%
  \BibitemOpen
  \bibfield  {author} {\bibinfo {author} {\bibfnamefont {M.}~\bibnamefont
  {Koashi}},\ }\href@noop {} {\enquote {\bibinfo {title} {Efficient quantum key
  distribution with practical sources and detectors},}\ } (\bibinfo {year}
  {2006}),\ \bibinfo {note} {preprint: quant-ph/0609180v1
  (unpublished)}\BibitemShut {NoStop}%
\bibitem [{\citenamefont {Rice}\ and\ \citenamefont
  {Harrington}(2009)}]{rice09}%
  \BibitemOpen
  \bibfield  {author} {\bibinfo {author} {\bibfnamefont {P.}~\bibnamefont
  {Rice}}\ and\ \bibinfo {author} {\bibfnamefont {J.}~\bibnamefont
  {Harrington}},\ }\href@noop {} {\enquote {\bibinfo {title} {Numerical
  analysis of decoy state quantum key distribution protocols},}\ } (\bibinfo
  {year} {2009}),\ \bibinfo {note} {preprint: quant-ph/0901.0013v2}\BibitemShut
  {NoStop}%
\bibitem [{\citenamefont {All\'{e}aume}\ \emph {et~al.}(2009)\citenamefont
  {All\'{e}aume}, \citenamefont {Roueff}, \citenamefont {Diamanti},\ and\
  \citenamefont {L\"{u}tkenhaus}}]{alleaume09}%
  \BibitemOpen
  \bibfield  {author} {\bibinfo {author} {\bibfnamefont {R.}~\bibnamefont
  {All\'{e}aume}}, \bibinfo {author} {\bibfnamefont {F.}~\bibnamefont
  {Roueff}}, \bibinfo {author} {\bibfnamefont {E.}~\bibnamefont {Diamanti}}, \
  and\ \bibinfo {author} {\bibfnamefont {N.}~\bibnamefont {L\"{u}tkenhaus}},\
  }\bibfield  {title} {\enquote {\bibinfo {title} {Topological optimization of
  quantum key distribution networks},}\ }\href@noop {} {\bibfield  {journal}
  {\bibinfo  {journal} {New J. Phys.}\ }\textbf {\bibinfo {volume} {11}},\
  \bibinfo {pages} {075002} (\bibinfo {year} {2009})}\BibitemShut {NoStop}%
\bibitem [{itu()}]{ituG984}%
  \BibitemOpen
  \href@noop {} {}\bibinfo {note} {ITU Recommendation G.984.6,
  \textit{Gigabit-capable passive optical networks (GPON): Reach extension},
  http://www.itu.int/rec/T-REC-G.984.6-200803-I/en.}\BibitemShut {Stop}%
\bibitem [{sma()}]{smartcity}%
  \BibitemOpen
  \href@noop {} {}\bibinfo {note} {For examples, see
  http://stratfordsmartcity.ca/2012/01/fibre-comes-home/;
  http://chattanoogagig.com/;
  http://www.wired.com/epicenter/2011/01/internet-meets-smart-grid/.}\BibitemShut
  {Stop}%
\bibitem [{Ten()}]{TenG}%
  \BibitemOpen
  \href@noop {} {}\bibinfo {note} {For an example of 10~Gb receiver
  sensitivity, see
  http://www.jdsu.com/ProductLiterature/jxp-01dxax1-xx0\_ds\_oc\_ae.pdf.}\BibitemShut
  {Stop}%
\bibitem [{10G()}]{10GbEthernet}%
  \BibitemOpen
  \href@noop {} {}\bibinfo {note} {IEEE Standard 802.3ae-2002, \uppercase{IEEE}
  802.3ae 10Gb/s Ethernet Task Force, see
  http://grouper.ieee.org/groups/802/3/ae/index.html.}\BibitemShut {Stop}%
\bibitem [{\citenamefont {Auyeung}\ and\ \citenamefont
  {Yariv}(1978)}]{auyeung78}%
  \BibitemOpen
  \bibfield  {author} {\bibinfo {author} {\bibfnamefont {J.}~\bibnamefont
  {Auyeung}}\ and\ \bibinfo {author} {\bibfnamefont {A.}~\bibnamefont
  {Yariv}},\ }\bibfield  {title} {\enquote {\bibinfo {title} {Spontaneous and
  stimulated \uppercase{R}aman scattering in long low loss fibers},}\
  }\href@noop {} {\bibfield  {journal} {\bibinfo  {journal} {IEEE J. Quant.
  Electron.}\ }\textbf {\bibinfo {volume} {14}},\ \bibinfo {pages} {347--352}
  (\bibinfo {year} {1978})}\BibitemShut {NoStop}%
\bibitem [{\citenamefont {Subacius}\ \emph {et~al.}(2005)\citenamefont
  {Subacius}, \citenamefont {Zavriyev},\ and\ \citenamefont
  {Trifonov}}]{subacius05}%
  \BibitemOpen
  \bibfield  {author} {\bibinfo {author} {\bibfnamefont {D.}~\bibnamefont
  {Subacius}}, \bibinfo {author} {\bibfnamefont {A.}~\bibnamefont {Zavriyev}},
  \ and\ \bibinfo {author} {\bibfnamefont {A.}~\bibnamefont {Trifonov}},\
  }\bibfield  {title} {\enquote {\bibinfo {title} {Backscattering limitation
  for fiber-optic quantum key distribution systems},}\ }\href@noop {}
  {\bibfield  {journal} {\bibinfo  {journal} {Appl. Phys. Lett.}\ }\textbf
  {\bibinfo {volume} {86}},\ \bibinfo {eid} {011103} (\bibinfo {year}
  {2005})}\BibitemShut {NoStop}%
\end{thebibliography}

%

\end{document}